# Winds and Disk Turbulence Exert Equal Torques on Thick Magnetically Arrested Disks

Vikram Manikantan[1,2,3], Nicholas Kaaz[2,3], Jonatan Jacquemin-Ide[2,3], Gibwa Musoke[4,5], Koushik Chatterjee[6],
Matthew Liska[7,8,9], and Alexander Tchekhovskoy[2,3]
[1] Steward Observatory & Department of Astronomy, University of Arizona, Tucson, AZ 85721, USA; vik@arizona.edu
[2] Department of Physics & Astronomy, Northwestern University, Evanston, IL 60202, USA
[3] Center for Interdisciplinary Exploration & Research in Astrophysics (CIERA), Physics & Astronomy, Northwestern University, Evanston, IL 60208, USA
[4] Anton Pannekoek Institute for Astronomy, University of Amsterdam, Science Park 904, 1098 XH Amsterdam, The Netherlands
[5] Canadian Institute for Theoretical Astrophysics, University of Toronto, 60 St. George Street, Toronto, ON M5S 3H8, Canada
[6] Black Hole Initiative at Harvard University, 20 Garden Street, Cambridge, MA 02138, USA
[7] Center for Relativistic Astrophysics, Georgia Institute of Technology, Howey Physics Bldg, 837 State St NW, Atlanta, GA 30332, USA
[8] Institute for Theory and Computation, Harvard University, 60 Garden Street, Cambridge, MA 02138, USA
Received 2023 October 25; revised 2024 February 29; accepted 2024 March 8; published 2024 April 18

## Abstract

The conventional accretion disk lore is that magnetized turbulence is the principal angular momentum transport process that drives accretion. However, when dynamically important large-scale magnetic fields thread an accretion disk, they can produce mass and angular momentum outflows, known as *winds*, that also drive accretion. Yet, the relative importance of turbulent and wind-driven angular momentum transport is still poorly understood. To probe this question, we analyze a long-duration ($1.2 \times 10^5 r_g/c$) simulation of a rapidly rotating ($a = 0.9$) black hole feeding from a thick ($H/r \sim 0.3$), adiabatic, magnetically arrested disk (MAD), whose dynamically important magnetic field regulates mass inflow and drives both uncollimated and collimated outflows (i.e., winds and jets, respectively). By carefully disentangling the various angular momentum transport processes within the system, we demonstrate the novel result that disk winds and disk turbulence both extract roughly equal amounts of angular momentum from the disk. We find cumulative angular momentum and mass accretion outflow rates of $\dot{L} \propto r^{0.9}$ and $\dot{M} \propto r^{0.4}$, respectively. This result suggests that understanding both turbulent and laminar stresses is key to understanding the evolution of systems where geometrically thick MADs can occur, such as the hard state of X-ray binaries, low-luminosity active galactic nuclei, some tidal disruption events, and possibly gamma-ray bursts.

*Unified Astronomy Thesaurus concepts:* Black holes (162); High energy astrophysics (739); Accretion (14); Relativistic jets (1390); Relativistic disks (1388)

## 1. Introduction

Black holes (BHs) are responsible for some of the most energetic phenomena in the Universe. This includes active galactic nuclei (AGN; Merloni et al. 2003; Blandford et al. 2019), gamma-ray bursts, tidal disruption events, and X-ray binaries (Mirabel & Rodríguez 1999; Corbel et al. 2000; Gallo et al. 2003, 2005). The emission generated by these systems is often accompanied by outflows such as jets that are relativistic, highly collimated, and BH powered (Blandford & Znajek 1977), or winds that are nonrelativistic, higher in gas density, and powered by the accretion disk (Blandford & Payne 1982). An accretion disk is an essential component of the central engine that powers these outflows. While the accretion disk's radiative signature varies from one system to the next, the inferred accretion rate is often highly correlated with the jet emission (Corbel et al. 2003; Markoff et al. 2003; Corbel et al. 2013; Zamaninasab et al. 2014). This correlation links accretion and outflows (also known as ejection) as interdependent dynamical processes. The key feature that links these processes is thought to be a magnetic field threading the accretion disk.

While not all systems launch jets, winds appear to be ubiquitous (Ponti et al. 2012; Cicone et al. 2014; Tetarenko et al. 2018; Mata Sánchez et al. 2023). The source of the wind and its connection to the jet and the accretion flow remains debated. In particular, it is unclear whether radiative or magnetic forces are primarily responsible for driving these winds (Chakravorty et al. 2016, 2023; Ponti et al. 2016; Tomaru et al. 2019, 2023; Fukumura et al. 2021; Waters et al. 2021).

A key piece of the puzzle is angular momentum transport, as angular momentum needs to be removed from the disk for it to accrete into the central BH. Accretion is thought to be driven by three different mechanisms: (a) via turbulent (magnetic and/ or hydrodynamic) small-scale torques, often thought to be due to the magnetorotational instability (MRI; Balbus & Hawley 1991), and usually modeled with an $\alpha$-parameter viscosity prescription (Shakura & Sunyaev 1973); (b) via large-scale poloidal magnetic fields anchored to the disk (Blandford & Payne 1982; Wardle & Koenigl 1993; Ferreira & Pelletier 1995); or (c) via nonaxisymmetric features such as spiral waves, eccentricity, or warps (e.g., Kaaz et al. 2023; Liska et al. 2023). In this work, we focus on the first two mechanisms.

Disk turbulence and wind outflows have historically been studied separately. However, they coexist in global simulations. Furthermore, steady-state accretion-ejection solutions that describe wind outflows require effective or nonideal magnetic resistivity, which can neatly be provided by MRI-driven turbulence within the accretion disk (Ferreira & Pelletier 1993). Linear modes of MRI can then saturate into nonlinear accretion-ejection structures (Lesur et al. 2013). Conversely, linear modes of MRI can emerge naturally in nonlinear accretion-ejection solutions (Jacquemin-Ide et al. 2019). Hence, both mechanisms—MRI and magnetohydrodynamic (MHD) wind launching—might be more deeply linked than previously thought.

---

[9] John Harvard Distinguished Science and ITC Fellow.

We note that both turbulent torques and large-scale outflow torques have distinct impacts on the secular evolution of the accretion disk. On the one hand, MRI-driven turbulence will lead to viscous spreading of the accretion disk, elongating the depletion timescale. On the other hand, outflow torques eject mass, which can lead to an increased mass-loss rate from the accretion disk. The ratio of the torques is thus an important parameter in secular evolution models (Tabone et al. 2022).

The magnetic field structure is also dependent on turbulent and large-scale outflow torques. The large-scale accretion flow will advect the magnetic field inward (Lubow et al. 1994; Rothstein & Lovelace 2008; Guilet & Ogilvie 2014). Turbulence not only generates a turbulent torque and viscously spreads the accretion disk, but also leads to turbulent diffusion of the large-scale magnetic field. Therefore, if the dominant torque is turbulent, the magnetic field would tend to dissipate outward on long timescales (Lubow et al. 1994). In contrast, large-scale magnetic torques that are associated with an outflow avoid contributing to the turbulent diffusion. This leads to efficient advection of the magnetic field (Rothstein & Lovelace 2008).

When accretion disks become fully saturated with large-scale poloidal magnetic fields that obstruct gas infall, we refer to them as magnetically arrested disks (MADs). The MAD is a robustly defined steady-state in general relativistic magnetohydrodynamic (GRMHD) simulations (Igumenshchev et al. 2003; Narayan et al. 2003; Tchekhovskoy et al. 2011). MADs are the natural end state of a system with a large reservoir of poloidal magnetic flux (e.g., field lines pointed in the radial and polar directions). Even when the initial magnetic flux is insufficient to lead to a MAD, the disk can eventually become a MAD if the mass accretion rate drops in time (Tchekhovskoy et al. 2014; Tchekhovskoy & Giannios 2015; Christie et al. 2019; Jacquemin-Ide et al. 2021; Gottlieb et al. 2023). Furthermore, they have historically been associated with so-called radiatively inefficient or advection-dominated accretion flows, although the paradigm has now broadened to include thinner radiative MADs (Avara et al. 2016; Morales-Teixeira et al. 2018; Liska et al. 2019, 2022a; Scepi et al. 2024). Moreover, if the host BH is spinning, they are accompanied by powerful relativistic jets. There is growing acceptance that many accretion disks in nature may be strongly magnetized, as this would address a variety of issues in both accretion theory and phenomenology. For instance, the classical Shakura & Sunyaev (1973) disk is thermally unstable in its inner regions that are dominated by radiation pressure (Lightman & Eardley 1974; Shakura & Sunyaev 1976; Piran 1978), and in the case of supermassive black hole (SMBH) disks, it is unstable to star formation in its outer regions. However, magnetically supported disks are both thermally stable (Begelman & Pringle 2007; Sądowski 2016) and stable to star formation (Gaburov et al. 2012). Additionally, observations of a rapid variability of months to years in AGN are incompatible with turbulent transport mechanisms in thin disks (Matt et al. 2003; Lawrence 2018), but they may be enabled by magnetized disks, which have systematically higher aspect ratios (Dexter & Begelman 2019) and strong outflows, both enabling rapid inflow. Recent Event Horizon Telescope (EHT) observations of both M87 (Akiyama 2019) and Sgr A* (EHT Collaboration et al. 2022) have also bolstered the expectation that many low-luminosity AGN in the Universe harbor MADs, as in both cases the observed horizon-scale disks are consistent with synthetic observations produced via ray-traced MAD simulations.

We also note that it has recently been called into question whether the MRI drives turbulence in MADs or if the MRI is quenched by the strong magnetic field and another instability, such as magnetic interchange or the Rayleigh–Taylor instability, instead drives turbulence (Kaisig et al. 1992; Lubow & Spruit 1995; Spruit et al. 1995; Marshall et al. 2018; Mishra et al. 2019). For instance, McKinney et al. (2012) and White et al. (2019) both found that in MADs, the wavelength of the marginally stable MRI mode is larger than the density scale height of the disk, suggesting that MADs are stable to MRI-driven turbulence. We note, however, that this analysis only considers the unstratified axisymmetric MRI and not the nonaxisymmetric MRI. The nonaxisymmetric MRI survives to much higher magnetic field strengths (Fragile & Sądowski 2017; Begelman et al. 2022) and could drive turbulence in MADs.

In this paper, we characterize the angular momentum transport within a geometrically thick MAD. We analyze the angular and radial dependence of angular momentum transport by decomposing the system into three regions: the disk, winds, and jets. We take inventory of the angular momentum transport processes by separating the turbulent and laminar components of both the hydrodynamic (Reynolds) and magnetic (Maxwell) stresses. In Section 2, we describe our simulation setup and diagnostics. In Section 3, we establish the angular and radial dependence of angular momentum transport in the system, and in Section 4 we present our conclusions.

## 2. Approach

### 2.1. Computational Method and Simulation Setup

We use the H-AMR code (Liska et al. 2022b), which evolves the equations of ideal GRMHD on a spherical polar grid in modified Kerr-Schild coordinates (Gammie et al. 2003) using the piecewise parabolic method for spatial reconstruction (Colella & Woodward 1984) and second-order time-stepping. H-AMR uses various computational speed-ups, including both static and adaptive mesh refinement (SMR and AMR, respectively), local adaptive time-stepping, and acceleration by graphical processing units (GPUs). We adopt dimensionless units $G = M = c = 1$, where $M$ is the mass of the BH, such that the gravitational radius $r_\mathrm{g} = GM/c^2 = 1$.

In this paper, we study a long-duration (out to $120,000 r_g/c$) simulation that was originally presented in Liska et al. (2020). We consider a rapidly rotating BH of dimensionless spin $a = 0.9$ at the center of an equilibrium Chakrabarti (1985) torus with a sub-Keplerian angular momentum profile, $l \propto \lambda^{1/4}$ (where $\lambda$ is the radius of the relativistic von Zeipel cylinder, which is asymptotically the cylindrical radius), with an inner edge located at $r_\mathrm{in} = 6 r_g$ and a pressure maximum at $r_\mathrm{max} = 13.792 r_g$. The outer edge of the torus, which is determined by our angular momentum distribution, lies at $r_\mathrm{out} \approx 4 \times 10^4 r_g$. We use a polytropic equation of state with an adiabatic index $\gamma = 5/3$, which results in a torus scale height that ranges from $H/r \sim 0.2$ at $r_\mathrm{max}$ to $H/r \sim 0.5$ at $r_\mathrm{out}$.

The outer edge of the grid is located at $r = 10^5 r_g$, and the inner edge is sufficiently far below the event horizon such that our inner boundary is causally disconnected from the BH exterior. The grid is uniformly spaced in $\log r$, $\theta$, and $\varphi$ for the ranges $0.1127 < \log r/r_g < 5$, $0 < \theta < \pi$, and $0 < \varphi < 2\pi$. We use transmissive boundary conditions (BCs) at the poles, periodic BCs in the $\varphi$-direction, and absorbing BCs at the inner and outer radial boundaries. We select a base resolution of $N_r \times N_\theta \times N_\varphi = 1872 \times 624 \times 128$ with four levels of static mesh refinement in $\varphi$. This means that the $\varphi$ resolution reaches 1024 cells at the midplane and decreases at higher latitudes until it reaches 128 cells near the pole. This results in a total of over one billion cells and 70–90 cells per disk scale height (taking $H/r \approx 0.4$).

Within the torus, we initialize a toroidal magnetic field with a (nearly) uniform plasma $\beta \equiv p_{\rm gas}/p_{\rm mag} = p_{\rm gas}/b^2 = 5$, where $p$ and $b^2$ are the fluid-frame gas and magnetic pressure, respectively. Via dynamo action, the initially toroidal magnetic field evolves into a large-scale poloidal magnetic field at $\sim$10,000$r_g/c$ and soon thereafter pushes the disk into the MAD regime (see Liska et al. 2020, and references therein for a full discussion).

### 2.2. Energy-momentum Tensor Decompositions

One of the main goals of this work is to take inventory of the angular momentum transport mechanisms in MADs. To measure the angular momentum flux, we use the mixed stress-energy-momentum tensor,

$$T^\mu_\nu = (\rho + u + p + b^2)u^\mu u_\nu + \left(p + \frac{1}{2}b^2\right)\delta^\mu_\nu - b^\mu b_\nu, \quad (1)$$

where $\rho$ is the gas density, $u$ is the internal energy, and $p$ is the gas pressure, each measured in the fluid-frame. $b^\mu$ is the contravariant magnetic field four-vector (such that $b^2 = b^\mu b_\mu$), $u^\mu$ is the contravariant four-velocity, and $\delta^\mu_\nu$ is the Kronecker delta. We also use the relativistic enthalpy, $w \equiv \rho + u + p + b^2$. In our decomposition, we separate the first term in Equation (1), which represents hydrodynamic stresses (except for a small contribution from the magnetic pressure),

$$(T^\mu_\nu)_u = wu^\mu u_\nu, \quad (2)$$

and the third term, which represents magnetic stresses,

$$(T^\mu_\nu)_{EM} = -b^\mu b_\nu. \quad (3)$$

Since we only are concerned with off-diagonal terms ($\mu \neq \nu$) of $T^\mu_\nu$ in this work, the second term in Equation (1) is always zero, and we do not consider it further. We average our stresses in space and time using the following procedure. First, we perform Reynolds decompositions for the various terms. If a stress equals the product $XY$, then we define the Reynolds decomposition as

$$\langle XY \rangle \equiv \langle X \rangle \langle Y \rangle + \langle \delta X \delta Y \rangle,$$

where we have defined our averages over space and time as

$$\langle X \rangle \equiv \iint \sqrt{-g} X \, dt \, d\varphi \,\Big/\, \iint \sqrt{-g} \, dt \, d\varphi. \quad (4)$$

Here, quantities of the form $\langle X \rangle \langle Y \rangle$ represent the laminar large-scale behavior of the flow, while quantities of the form $\langle \delta X \delta Y \rangle$ represent the turbulent small-scale behavior of the flow. Our time averages are taken over the period 80,000–120,000$r_g/c$. When calculating the turbulent $\delta X$ term, we first calculate $\langle X \rangle$ on the entire data set, then iterate through each time step to find the difference between the instantaneous data and the averaged data (i.e., $\delta X = X - \langle X \rangle$).

We perform our Reynold's decomposition on Equations (2) and (3) to derive the laminar and turbulent components of the hydrodynamic and magnetic stresses, respectively. The decomposition of Equation (2) is

$$(T^\mu_\nu)_{\delta u} = \langle (wu^\mu - \langle wu^\mu \rangle)(u_\nu - \langle u_\nu \rangle) \rangle = \langle \delta(wu^\mu) \delta u_\nu \rangle, \quad (5)$$

$$(T^\mu_\nu)_{\langle u \rangle} = \langle wu^\mu \rangle \langle u_\nu \rangle. \quad (6)$$

Moving forward, we refer to Equation (5) as the turbulent Reynolds stress and to Equation (6) as the advective flow of angular momentum. We decompose Equation (3) in a similar fashion,

$$(T^\mu_\nu)_{\delta b} = -\langle (b^\mu - \langle b^\mu \rangle)(b_\nu - \langle b_\nu \rangle) \rangle = -\langle \delta b^\mu \delta b_\nu \rangle, \quad (7)$$

$$(T^\mu_\nu)_{\langle b \rangle} = -\langle b^\mu \rangle \langle b_\nu \rangle. \quad (8)$$

We refer to Equation (7) as the turbulent Maxwell stress and to Equation (8) as the large-scale Maxwell stress. We return to these calculations when we present them in Section 3.3.

## 3. Results

To study the steady-state angular momentum transport mechanisms in our accretion disk, we time- and $\varphi$-average all quantities in our simulation for 80,000–120,000$r_g/c$ with a cadence of $10r_g/c$, providing a total of 4001 snapshots. This corresponds to $\sim$1000 Keplerian orbits at $10r_g$ and $\sim$40 orbits at $100r_g$. The time-averaged mass accretion rate is constant up to $\sim$80 $r_g$, indicating that the system has reached steady state out to to this distance, which we set as the outer radius of our analysis. At this late time in the simulation, the accretion disk has also reached the MAD state with, on average, a constant accretion rate and magnetic flux on the event horizon (for a detailed description of the MAD state, see Tchekhovskoy et al. 2011).

### 3.1. Instantaneous Features of the Accretion Flow

Time- and $\varphi$-averages are necessary to understand the long-term behavior of MADs. This, however, smooths over many interesting transient structures in the accretion flow that we would like to highlight first.

Figure 1 depicts the turbulent nature of the accretion disk by plotting the plasma $\beta$ of the fluid at $t = 120{,}000\ r_g/c$. In Figure 1(a), we show a slice of plasma $\beta$ in the $x$–$z$ plane. Within the equatorial accretion disk, we see regions that are dominated by gas pressure (blue) and by magnetic pressure (red). These regions are turbulent and irregular, except for one large-scale vertically extended structure at $x \sim -50\ r_g/c$. This is a magnetic flux eruption event, which is a standard feature of MADs (Tchekhovskoy et al. 2011). Magnetic flux eruptions travel radially outward through the accretion disk and can be identified as large-scale magnetic flux bundles characterized by relatively lower values of $\beta$. Chatterjee & Narayan (2022) suggest that these flux eruption events provide an important form of angular momentum transport in MADs.

In Figure 1(b), we show an equatorial slice through the system. We now observe both small-scale turbulence and highly magnetized spiral modes, which is a standard feature of MADs and arises from the magnetic interchange instability (Kaisig et al. 1992; Lubow & Spruit 1995; Spruit et al. 1995; Marshall et al. 2018; Mishra et al. 2019). Indeed, we see the $x$–$y$ projection of the magnetic flux eruption event in Figure 1(a) at $x \sim -50r_g$, $y \sim -10r_g$.

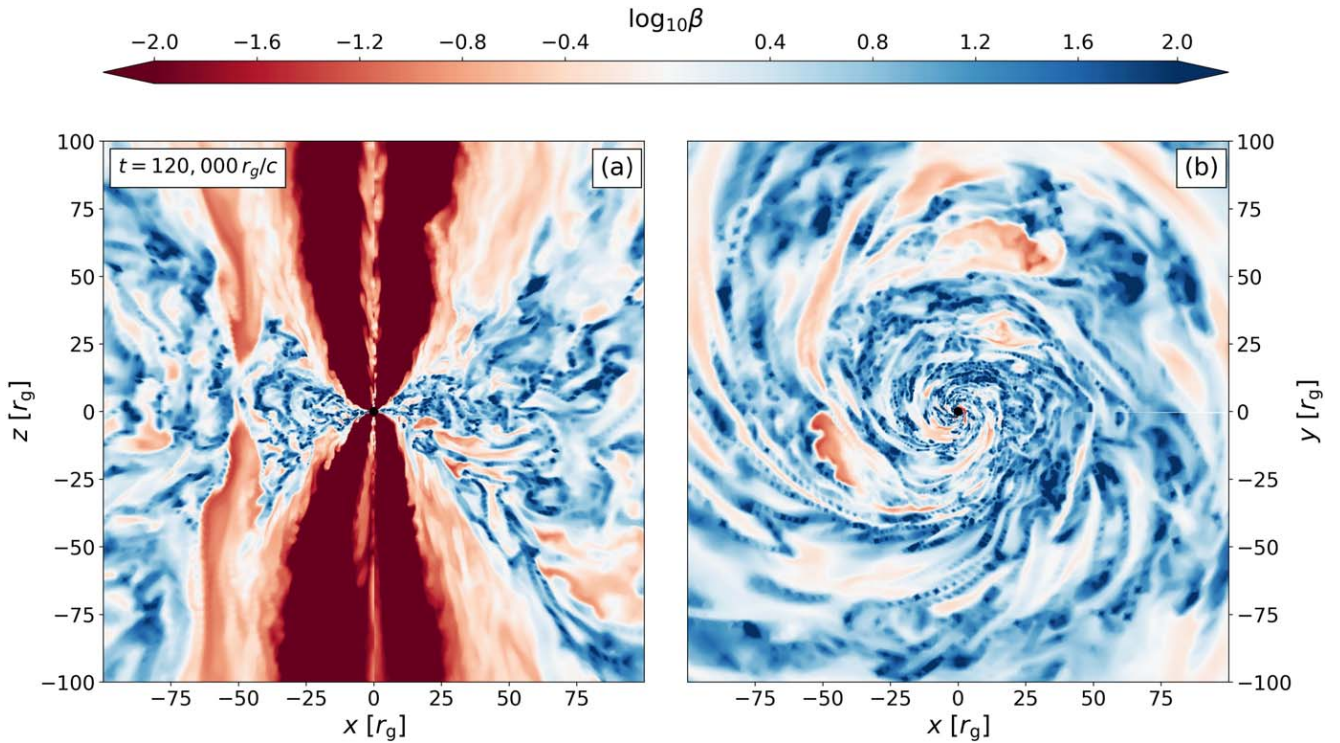

**Figure 1.** The turbulent nature of the instantaneous accretion disk makes obtaining a time-independent understanding of the system difficult. We depict an instantaneous snapshot of the system out to $100r_g$ in the ($x$–$z$), panel (a), and ($x$–$y$), panel (b), planes with a color map of the plasma $\beta$ parameter, where high (low) values of $\beta$ indicate less (more) strongly magnetized plasma. We note a pair of vertically elongated magnetic flux bundles (left side of panel (a), $x \sim -50r_g$) that radially escape from the event horizon.

### 3.2. Structure of the Angular Momentum Flow

In this section, we study the angular momentum flow in three regions of our system: in the accretion disk, in the disk winds, and in jets. We define each region using our time-averaged axisymmetrized data. Then, we plot these definitions in Figure 2, which shows $x$-$z$ slices of the flow that we $\varphi$- and time-averaged from 80,000 $r_g/c$ to 120,000 $r_g/c$, where the accretion flow has reached a quasi-steady state.

Figures 2(a) and (b) show the gas density $\rho$, and panels (c) and (d) show the plasma $\beta$. At the origin of each panel [$x = 0$, $z = 0$], we have plotted the BH event horizon. The left column (Figures 2(a), (c)) zooms in on the right column (Figures 2(b), (d)) from $100r_g$ to $25r_g$. We plot streamlines over the density and plasma $\beta$ contours to show the flow of the angular momentum flux, as defined by the four-vector $T^\mu_\varphi$ projected onto the Cartesian basis vectors $\hat{x}$ and $\hat{z}$. We use these streamlines to define the three regions of our system: jets, winds, and the accretion disk.

The jet regions exist at the poles of the BH and form an hourglass shape (at $x = y = 0$ and extending vertically in $z$). Here, the magnetic pressure increases to over 100 times the gas pressure (e.g., $\log_{10}\beta \lesssim -2$)[10] as the jet is magnetically launched and angular momentum is extracted from the BH through the Blandford–Znajek (BZ; Blandford & Znajek 1977) mechanism. For each of the two jets, we define their boundaries as the last radially outward angular momentum streamline (counting from the pole) to originate from the BH event horizon. Figure 2 shows the jet boundaries with dashed black and white lines. We adopt this robust definition as it conserves the radial time-averaged angular momentum flux within the jet. We also note that there are angular momentum streamlines anchored to the accretion disk within the BH ergosphere. However, the ergosphere lies within the innermost stable circular orbit (ISCO) for our choice of BH spin ($a = 0.9$) (Bardeen et al. 1972; Carroll 2019). Therefore, the disk within the ergosphere has a negligible specific angular momentum compared to the magnitude of the angular momentum extracted within the jet and from the disk outside of the ergosphere. Additionally, any angular momentum transport measured would be measured as a Blandford–Payne (i.e., wind) outflow, just as intended.

---

[10] We note that the value of $\beta$ in the jet region is limited by our prescribed ceiling for the magnetization of the flow, $\sigma = b^2/\rho < 20$.

We define all other radially outward angular momentum flux streamlines ($T^r_\varphi$), which are anchored to the disk, as part of the winds. We define the disk-wind boundary by the set of turn-around points of the angular momentum streamlines, e.g., where the radial angular momentum flux vanishes ($T^r_\varphi$). We show this boundary in Figure 2 with solid green lines. This is a somewhat novel way of describing the vertical extent of the accretion disk. In the innermost regions of the disk, $r \lesssim (10–20)r_g$, the wind is squeezed between the jet and the disk and subtends a smaller solid angle as compared to at larger radii (e.g., $r \gtrsim 50\,r_g$). We also note that the surfaces $T^r_\varphi = 0$ and $\rho v^r = 0$ (turning point of the accretion flow) are not identical. At $r < 20r_g$, the difference is negligible. However, at $20\,r_g < r < 100r_g$, the vertical height of the surfaces as measured from the $x$–$y$ plane differs by up to $\sim 4r_g$, which is a difference of about a 5%. Using mass flux surfaces would likely produce the same results.

We define the accretion disk as the region where angular momentum streamlines have a negative radial component ($T^r_\varphi < 0$). The disk resides roughly within $z \lesssim (0.4$–$0.5)r$, where the gas density peaks at $\rho \sim 10^2$, but is not significantly denser than the wind region.

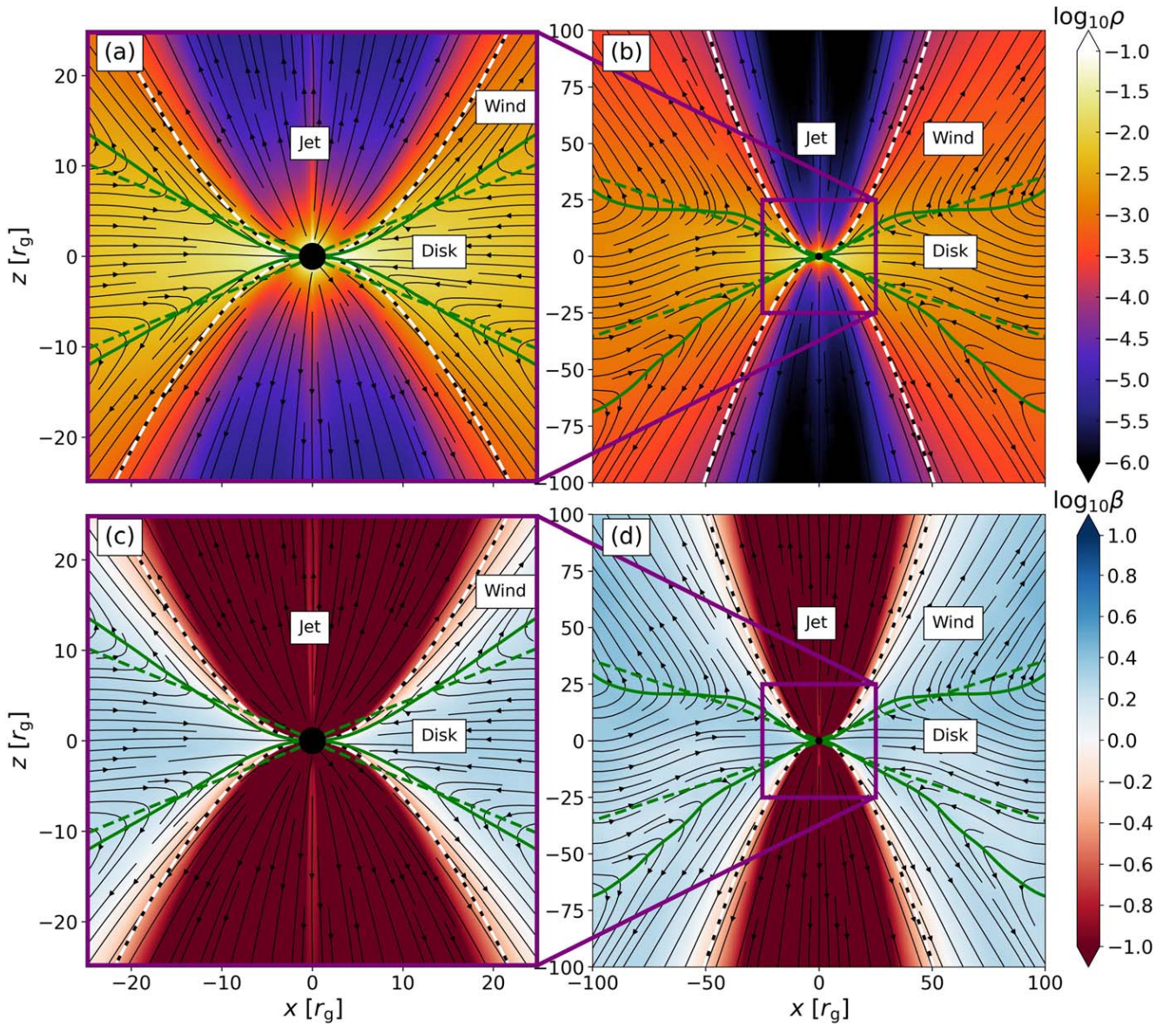

**Figure 2.** On average, MADs have a well-defined hourglass-shaped structure, consisting of an inflowing disk region and outflowing wind and jet regions. We depict the time-averaged axisymmetrized flow of angular momentum within the disk, wind, and jet regions at time $t = 80,000–120,000 \, r_g/c$ out to $100 \, r_g$ in the right column and out to $25 \, r_g$ in the left column. In the top row, we plot contours of the gas density, and in the bottom row, we plot contours of the plasma $\beta$. We have labeled each region: The solid green lines are the disk-wind boundaries, and the solid black and white lines are the wind-jet boundaries. The dashed green line shows the typical thermal definition of the disk thickness (Equation (9)). The black streamlines sketch the flow of angular momentum (Equation (1)).

The typical definition of the disk thickness is set by the midplane temperature of the accretion disk,

$$\frac{H}{r} \sim \frac{c_s}{v_{\rm rot}}, \quad (9)$$

where $c_s$ is the sound speed, and $v_{\rm rot}/r = u^\varphi/u^t$, and $u^\mu$ is the four-velocity of the fluid. We plot this definition in Figure 2 with dashed green lines. We can see that it underpredicts the disk thickness for $r \lesssim 20 r_g$ and overpredicts it for $r \gtrsim 20 r_g$ compared to our streamline-based definition. Furthermore, this definition assumes vertical hydrostatic equilbrium, which is not satisfied near the BH as the disk is squeezed vertically between the two jets. Thus, we opt for our definition instead as it is specifically constructed to separate regions of angular momentum inflow and outflow.

Figures 2(b) and (d) show that the angular momentum flow becomes asymmetric across the midplane for large radii ($r > 50 \, r_g$). While this may seem surprising, this asymmetry is not unprecedented, as in totally energy-conserving MADs the turbulent eddies have length scales $\mathcal{O}(H)$ that are $\mathcal{O}(r)$ since $H/r \lesssim 0.4$–$0.5$. These eddies can disrupt the current sheet at the midplane of the disk and cause asymmetries over long timescales (e.g., Tchekhovskoy 2015; Chatterjee & Narayan 2022).

### 3.3. Vertical Structure of Outflows: Disk versus Winds

In this section, we discuss the vertical profile of the angular momentum transport in our simulation. Here, we use the Reynolds decomposition of the stress-energy tensor (given by

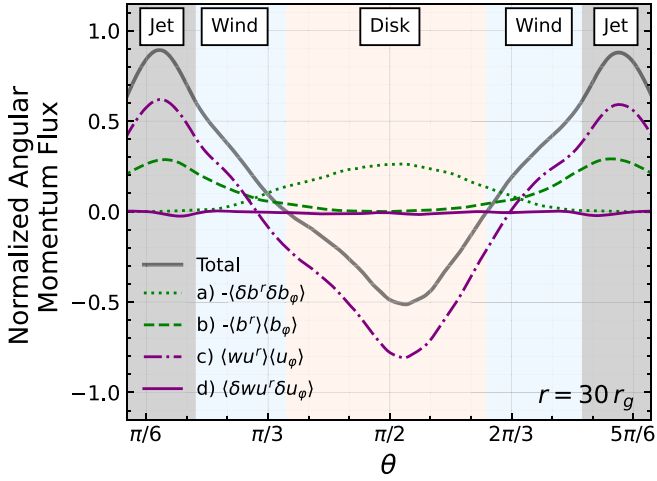

**Figure 3.** The relative importance of each angular momentum transport mechanism strongly depends on the latitude. We depict the polar angular dependence of the decomposed time-averaged angular momentum flux at $r = 30 r_g$, normalized to the average angular momentum flux at $r = 2 r_g$ (Equation (10)). We break down the total radially outward angular momentum flux ($T^r_\varphi$; Equation (1)) into four components: (a) turbulent Maxwell stresses ($-\langle \delta b^r \delta b_\varphi \rangle$; Equation (7)), (b) large-scale laminar Maxwell stresses ($-\langle b^r \rangle \langle b_\varphi \rangle$; Equation (8)), (c) advective transport ($\langle w u^r \rangle \langle u_\varphi \rangle$; Equation (6)), and (d) turbulent Reynolds stresses ($\langle \delta w u^r \delta u_\varphi \rangle$; Equation (5)). The turbulent Maxwell stresses are the dominant mode of outward angular momentum transport within the disk, while the large-scale Maxwell and advective terms dominate in the winds.

Equations (5)–(8)). Figure 3 shows the terms contributing to the radial angular momentum transport as a function of $\theta$ at $r = 30 r_g$. Positive values indicate radially outward flow. We have normalized the y-axis to the average angular momentum flux onto a sphere at $2 r_g$,

$$(G^r_\varphi)_{2 r_g} = 2\pi \left[ \int \langle T^r_\varphi \rangle \sqrt{-g} \, d\theta \right]_{r = 2 r_g}. \quad (10)$$

Here, the factor of $2\pi$ results from a trivial $\varphi$ integral over the axisymmetrized fluxes. The streamlines in Figure 2 show the total angular momentum flux, which is also given by the solid gray line in Figure 3. It is directed outward in the wind regions and inward in the accretion disk. The advective transport ($\langle w u^r \rangle \langle u_\varphi \rangle$; Equation (6)) is radially outward in the winds and inward in the accretion disk. The Reynolds stresses ($\langle \delta w u^r \delta u_\varphi \rangle$; Equation (5)) are negligible in comparison to the other modes of transport. We break down the radial angular momentum transport driven by Maxwell stresses (Equations (7) and (8)) into turbulent, $-\langle \delta b^r \delta b_\varphi \rangle$, and laminar, $-\langle b^r \rangle \langle b_\varphi \rangle$, components. Within the disk, the turbulent Maxwell stresses dominate the outward angular momentum transport. As we cross into the wind regions, the large-scale magnetically driven angular momentum transport begins to dominate (e.g., Blandford & Payne 1982). This is consistent with other simulations of highly magnetized disks (Jacquemin-Ide et al. 2021; Scepi et al. 2024).

In Figure 4(a), we present the radially outward transport of angular momentum within the three regions—the jets, the winds, and the disk—as a function of radius. We decompose the radial angular momentum fluxes in the wind and disk regions into their turbulent and laminar Maxwell and Reynolds stress components (see Equation (11)). Then, we integrate over the radially oriented surfaces of the disk and the wind regions,

$$(G^r_\varphi)_{\text{region}} = 2\pi \int_{\theta_{\min}(r)}^{\theta_{\max}(r)} \langle T^r_\varphi \rangle \sqrt{-g} \, d\theta, \quad (11)$$

where $\theta_{\min}$ and $\theta_{\max}$ are the minimum and maximum $\theta$ boundaries of the disk or wind regions, respectively. We ignore the turbulent Reynolds stress as it is insignificant compared to the other stresses (see Figure 3). On the x-axis, we plot the radius out to $100 r_g$, and on the y-axis, we plot the normalized angular momentum flux. We normalize the flux to the angular momentum flux through the sphere at $2 r_g$ (Equation (10)).

The horizontal gray line shows that the total angular momentum flux within the jet is constant (($G^r_\varphi)_{\text{jet}} = $ const.), which is by construction. The solid blue and red lines describe the total angular momentum fluxes in the wind and disk regions, respectively. The transport of angular momentum in the wind and by the disk stresses are in equipartition at all radii. Furthermore, we see that the total angular momenta transported by the disk and wind regions both follow the same power law in radius, $\propto r^{0.9}$. It is unclear which physical processes result in this power-law dependence. We link this power law to the mass ejection rate in Section 3.4.

We plot the various mechanisms of angular momentum transport within each region with the dashed (turbulent Maxwell stress; $-\langle \delta b^r \delta b_\varphi \rangle$), dotted (large-scale Maxwell stress; $-\langle b^r \rangle \langle b_\varphi \rangle$), and dash–dotted (advective flow; $-\langle u^r \rangle \langle u_\varphi \rangle$) lines. We break down the radially outward angular momentum flux within the disk into the turbulent Maxwell and large-scale Maxwell stresses. We do not include the inward radial advective term within the disk fluxes as it describes the net radially inward accretion due to outward vertical or radial angular momentum transport processes. We see that the turbulent Maxwell stress ($\langle \delta b^r \delta b_\varphi \rangle$) closely follows the extracted total angular momentum and is the dominant mode of angular momentum transport within the disk. The large-scale Maxwell stress ($\langle b^r \rangle \langle b_\varphi \rangle$) is highly subdominant within the disk, which is as expected because the disk is turbulent. We also note that since the plasma $\beta$ in the accretion disk is roughly constant out to $\sim 80 r_g$, we expect the relative strength of the magnetic and velocity stress-energy terms to remain constant. In disks that are only dominated by magnetic pressure in the inner regions (i.e., the plasma $\beta$ increases with radius), we do not expect our radial $\dot{L}$ profiles to hold. A disk like that might be better described by a broken power law where the inner magnetically saturated regions are described by our $\dot{L} \propto r^{0.9}$ profile, and the outer regions are not.

We break down the angular momentum transport within the winds into the turbulent Maxwell, large-scale Maxwell, and advective terms. For $r \lesssim 20 r_g$, the large-scale Maxwell stresses are the dominant mechanism for radially outward angular momentum transport. For $20 r_g \lesssim r \lesssim 70 r_g$, the advective term dominates. The transition between these transport mechanisms could be due to the acceleration of the wind by the magnetic forces, effectively redistributing the angular momentum transport from magnetic to advective transport. The advective component could also be driven by different forces, such as an enthalpy gradient, which would be naturally present in our hot total energy-conserving MAD. Below, we try to disentangle these two possibilities by looking at the angular momentum fluxes throughout the disk surface.

At distances $r > 70 r_g$, the turbulent Maxwell angular momentum flux exceeds the large-scale Maxwell angular momentum flux. However, at these radii and beyond, we only sample ≲10 orbits in our averages, and the disk might not have achieved quasi-steady state at these large radii. This may not be enough time to average over all variability timescales, and our averages at these radii must therefore be taken with a grain of salt.

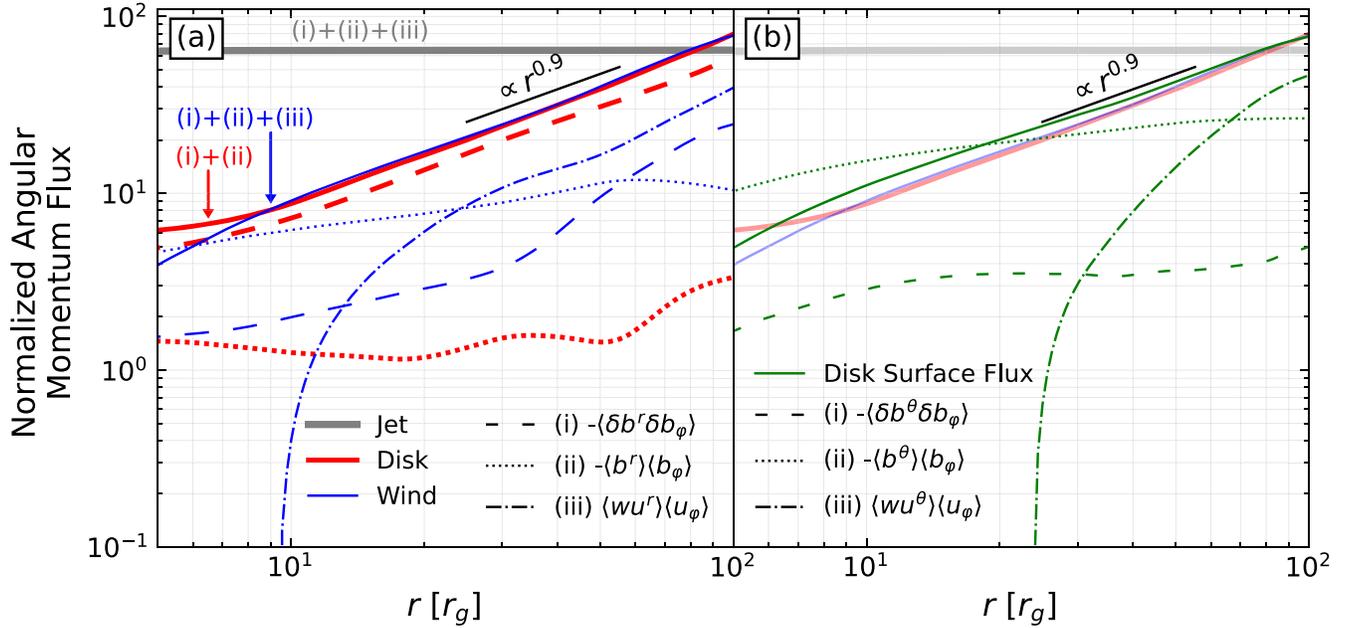

**Figure 4.** At all radii, the magnetized turbulence in the disk and laminar wind torques contribute almost equally to the radial transport of angular momentum. Here, we show the radial profile of the time- and $\varphi$-averaged decompositions of the angular momentum flux in our system. Panel (a): For each of our three main regions, we integrate $T_\varphi^r$ across the $\theta$-direction. The jet (thick solid gray line) carries a constant angular momentum outward by construction (Section 3.2). The accretion disk (thick solid red line) transports angular momentum outward with a magnitude comparable to the winds (solid blue line), and both are proportional to $r^{0.9}$. Note that we do not include the advective term ($\langle wu^r\rangle\langle u_\varphi\rangle$; Equation (6)) in the total disk angular momentum flux because it is negative and feeds the BH. Panel (b): The cumulative integral in radius of the angular momentum flux through the surface of the accretion disk (solid green line; see Equation (12)) has a similar magnitude as the wind and disk regions of panel (a) with the same $r^{0.9}$ dependence. For both, we further break down the angular momentum fluxes into (i) turbulent Maxwell stresses ($\langle \delta b^\mu \delta b_\varphi\rangle$; Equation (7)) with the dashed lines, (ii) large-scale laminar Maxwell stresses ($\langle b^\mu\rangle\langle b_\varphi\rangle$; Equation (8)) with the dotted lines, and (iii) advective transport ($\langle wu^\mu\rangle\langle u_\varphi\rangle$; Equation (6)) with the dash–dotted lines, where $\mu = r, \theta$. The turbulent Reynolds term is negligible everywhere (see the solid purple line in Figure 3). The dashed, dotted, and dash–dotted lines sum to their respective totals described by the solid red, blue, and green lines. The angular momentum outflows within the accretion disk, in the winds, and throughout the disk surface are comparable in magnitude at all radii, and they exceed the jet-powered outflows for $r > 80 r_g$.

In Figure 4(b), we show the cumulative integral in $r$ of the angular momentum flux through the disk-wind boundary and its decomposition, analogous to panel (a). Namely, here we decompose the angular momentum flux in the $\theta$ direction, $T_\varphi^\theta$, rather than $T_\varphi^r$. Our specific steps are as follows: we perform our decomposition and averaging procedure outlined in Section 2.2, we find the values of our decomposed $\langle T_\varphi^\theta\rangle$ at the north and south disk-wind boundaries, and we cumulatively integrate them in radius. We then sum the result from the top and bottom surfaces of the disk,

$$G_\varphi^\theta = 2\pi \int_{r_h}^r \sqrt{-g}\ \langle T_\varphi^\theta\rangle (r', \theta = \theta_{\text{bottom}})\ dr' \\ - 2\pi \int_{r_h}^r \sqrt{-g}\ \langle T_\varphi^\theta\rangle (r', \theta = \theta_{\text{top}})\ dr', \quad (12)$$

where $G_\varphi^\theta$ is the cumulative angular momentum flux through the surface of the accretion disk, $\theta_{\text{bottom}}$ and $\theta_{\text{top}}$ are the bottom and top disk-wind boundaries (solid green lines in Figure 2), and $r_h$ is the event horizon of the BH. As in Equations (10) and (11), the factor of $2\pi$ arises from a trivial $\varphi$ integral over the axisymmetrized fluxes. The flux through the top of the accretion disk is in the negative $\theta$ direction, and therefore, we place a negative sign in front of the $\theta_{\text{top}}$ term to find the net flow of angular momentum out of the accretion disk.

By definition, no angular momentum crosses the wind-jet boundary on average, and the divergence theorem tells us that in steady state, this integral, Equation (12), evaluated at radius $r$, equals the surface integral of $T_\varphi^r$ over the wind region at the same radius. Indeed, Figure 4(b) shows that the total disk surface flux (solid green line) is almost equal to the wind flux (solid light blue line) and follows the same $\propto r^{0.9}$ radial scaling. The slight discrepancy between the two is likely due to the wind-jet boundary close to the BH. Close to the BH, the wind and disk are squeezed by the strong magnetic field attached to the BH. In the inner regions, the winds become a thin vertical sheath that could easily be contaminated by the angular momentum being extracted from the BH. This discrepancy may be further exacerbated by the density floors applied to the low-density regions at the base of the jet, close to the BH. This artificially adds mass, and therefore, momentum, to the inner jet and wind regions. We see this uptick in the total mass accretion rate of the system, which we depict with a thick solid black line in Figure 5. Nonetheless, we stress that the agreement between the radial wind flux and the vertical disk surface flux is remarkable.

We decompose the angular momentum flux into its components at the disk surface (as detailed in Figure 4(b)) and show that the large-scale Maxwell torques, $\langle b^\theta\rangle\langle b_\varphi\rangle$, dominate the angular momentum transport for $r < 20\ r_g$, beyond which the advective term, $\langle wu^\theta\rangle\langle u_\varphi\rangle$, takes over.[11]

---
[11] The advective torque is negative (not shown) very close to the BH, $r < 10 r_g$. We believe that this is because the wind is unable to escape at very small radii due to it and the disk being squeezed by the strong magnetic field attached to the BH.

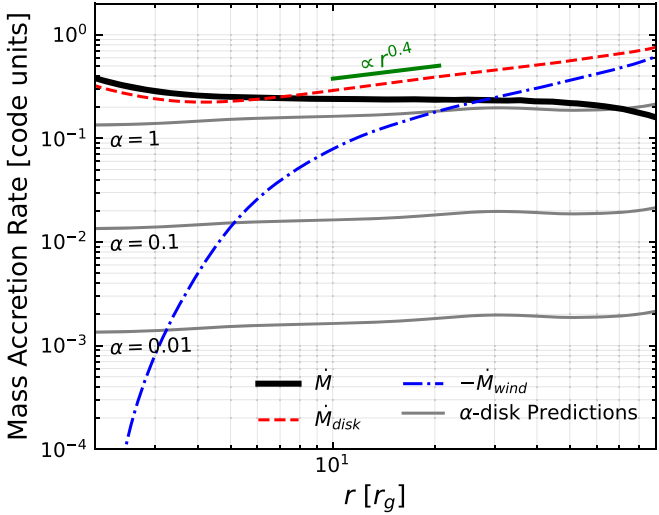

**Figure 5.** The mass accretion rate in our disk is comparable to that of an $\alpha$-theory with vigorous turbulence (e.g., $\alpha \sim 1$). Here, we compare the radial dependence of the mass accretion rate in the wind ($\dot{M}_{\text{wind}}$; dash–dotted blue line), the disk ($\dot{M}_{\text{disk}}$; dashed red line), and in the total system ($\dot{M}$; thick solid black line), with an analytical prediction for the mass accretion rate given three values for the $\alpha$ prescription ($\dot{M}_{\text{SS}}$; thin solid gray lines). We can see that an $\alpha \sim 1$ analytic prediction is closest to the actual mass accretion rate within the system. $\dot{M}_{\text{disk}}$ roughly obeys a power law with radius $\sim r^{0.4}$. Since the angular momentum transport obeys a $\sim r^{0.9}$ power law (Figure 4), this is consistent with a (sub-)Keplerian specific angular momentum profile $\sim r^{0.5}$.

To make these comparisons between flux components, we compare the gradient of each component because this describes the instantaneous strength of each flux at a radius (see Figure 6, Appendix). Note that the large-scale Maxwell torques remain dominant up to a similar radius when compared to panel (a).

In panel (b), we show that the advective torque, $\langle wu^\theta \rangle \langle u_\varphi \rangle$, becomes dynamically important in the outer regions of the accretion disk, $r > 20 r_g$. The advective torque is related to the vertical gradient of angular momentum and should be negligible for negligible vertical gradients of $\langle u_\varphi \rangle$ (see the fourth term of Equation (17) of Zhu & Stone (2018)). However, radiatively inefficient accretion disks such as the one modeled here can have sharp vertical angular momentum gradients related to their high thermal pressure. These sharp gradients can then lead to an efficient advective torque, $\langle wu^\theta \rangle \langle u_\varphi \rangle$, because gas that is displaced in the $+\hat{z}$ direction will carry its angular momentum into a region with lower angular momentum. This is in contrast with Scepi et al. (2024), who find that this term is always negligible even for their thicker disks. We emphasize that Scepi et al. (2024) employed explicit cooling to establish a thick disk, whereas our approach involves evolving the disk without any cooling. Despite yielding a similar geometric thickness, these methods may not produce identical dynamics, potentially influencing the wind launching (Casse & Ferreira 2000).

It is clear that the advective ($\langle wu^\theta \rangle \langle u_\varphi \rangle$) and laminar magnetic ($\langle b^\theta \rangle \langle b_\varphi \rangle$) torques both contribute to the total surface angular momentum transport. Hence, the winds in geometrically thick MADs should be understood as magneto-thermal (Casse & Ferreira 2000) and not purely driven by the Blandford & Payne (1982) mechanism. This is not surprising as the thermal energy of a geometrically thick MAD is considerable.

Furthermore, Scepi et al. (2024) find that the large-scale vertical torque, $\langle b^\theta \rangle \langle b_\varphi \rangle$, dominates the angular momentum transport in the inner regions of the accretion disk, $r < 20 r_g$, which we confirm with our findings in this study. However, unlike Scepi et al. (2024), our analysis extends to $20\, r_g < r \lesssim 70 r_g$, where we find that the turbulent radial Maxwell torque within the accretion disk, $\langle \delta b^r \delta b_\varphi \rangle$ (Figure 4(a)), is in equipartition with the total vertical torque at the disk surface. Last, we note that the vertical turbulent Maxwell term, $\langle \delta b^\theta \delta b_\varphi \rangle$, is subdominant throughout.

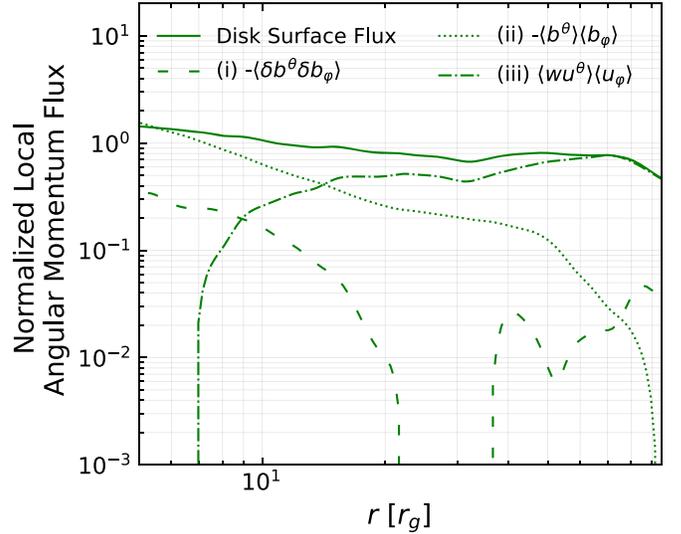

**Figure 6.** We plot the angular momentum flux through the disk surface, analogously to Figure 4(b), but instead of depicting the cumulative integral form of the angular momentum surface flux (Equation (12)), we instead depict the local angular momentum surface flux, $\frac{d}{dr} G_\varphi^\theta$ (see the Appendix). We further break down the local surface angular momentum flux into (i) turbulent Maxwell stresses ($\langle \delta b^\mu \delta b_\varphi \rangle$; Equation (7)), (ii) large-scale laminar Maxwell stresses ($\langle bu^\mu \rangle \langle b_\varphi \rangle$; Equation (8)), and (iii) advective transport ($\langle wu^\mu \rangle \langle u_\varphi \rangle$; Equation (6)), where $\mu = r, \theta$. The turbulent Reynolds term is negligible everywhere, and as in Figure 4, we therefore omit it. The dashed, dotted, and dash–dotted lines sum the total local angular momentum flux through the disk surface, denoted by the solid green line.

In panel (b) of Figure 4, we plot the cumulative integral of the surface fluxes along the disk-wind boundaries. This was done to properly conserve the angular momentum fluxes through our Gaussian surfaces. However, this can make it difficult to assess by eye the radial dependence of the various terms contributing to $G_\varphi^\theta$. To remedy this, we have included Figure 6 in the Appendix, where we depict $\frac{d}{dr} G_\varphi^\theta$ as a function of radius. This shows that the advective surface flux ($\langle wu^\theta \rangle \langle u_\varphi \rangle$) locally dominates the vertical angular momentum transport for $20 r_g < r \lesssim 70 r_g$, even though they only become globally dominant for $r > 70 r_g$ (see Figure 4(b)). We refer to the Appendix for further discussion.

### 3.4. Inflow Rates

In this section, we assess how efficiently our system can drive mass inflow. To do this, we compare the measured mass accretion rate in the winds ($\dot{M}_{\text{wind}}$) and the disk ($\dot{M}_{\text{disk}}$), and in the system as a whole ($\dot{M}$) to what classical $\alpha$ viscosity theory predicts (Shakura & Sunyaev 1973; Pringle 1981). The $\alpha$ prescription predicts the following mass accretion rate:

$$\dot{M}_{\text{SS}} = \alpha \frac{4\pi H P_{\text{gas}}}{\Omega}. \qquad (13)$$

Here, $H(r)$ is the vertical scale height of the accretion disk, and $\Omega(r)$ is the orbital frequency of the gas, which we have measured in the frame of the the zero angular momentum observer (ZAMO; see, e.g., definitions in Takahashi 2007). This frame is chosen to eliminate rotation due to frame-dragging by the central BH. $P(r)$ is the gas pressure in the disk. We evaluate the quantities $\Omega(r)$ and $P(r)$ using density-weighted averages of the solid angle, i.e., for instance, $P(r) = \int \langle p \rangle \langle \rho \rangle \sqrt{-g}\, dA_\theta / \int \langle \rho \rangle \sqrt{-g}\, dA_\theta$.

We can also define a mass ejection index (Ferreira & Pelletier 1995) to quantify the radial dependence of the mass accretion rate,

$$\xi = d\ln \dot{M}_{\rm disk}/d\ln r, \quad (14)$$

where $\dot{M}_{\rm disk}$ is the mass accretion rate within the disk, and $r$ is the radial coordinate. We then use Equation (13) to compute $\dot{M}_{\rm SS}$ for a range of values, $\alpha = \{0.01, 0.1, 1\}$, and we show the resulting radial profiles of $\dot{M}_{\rm SS}$ in Figure 5 with the solid gray lines. This allows us to compare our simulated mass accretion rate, shown with the thick black line, with the prediction of the $\alpha$ disk. First, we see that the total time-averaged mass accretion rate (thick solid black line) is roughly constant out to $\sim 80 r_g$, with a deviation of $\sim 10\%$. This is despite the asymmetries in the accretion flow in Figure 2, which suggests that the system has reached inflow equilibrium out to that radius. We also note that the uptick in $\dot{M}$ at $r \lesssim 5 r_g$ is due to the density floors, which add mass in the jets. We then separate the mass flux into the wind and disk components using the definitions of Section 3.2.

The mass accretion rates in the winds (dash–dotted blue line) and disk (dashed red line), however, are not constant because the disk is losing gas to the wind. This results in a roughly $\propto r^{0.4}$ radial dependence of the accretion rate in the disk. Due to mass conservation, the wind should exhibit the same power law. This power law is altered at small radii, however, due to the boundary condition on the accretion rate in each region: The disk accretion rate must approach the BH accretion rate, while the wind accretion rate must approach zero. In other words, in the inner regions ($r \gtrsim r_{\rm isco}$), the wind region carries almost no mass (see Figure 5). In the outer regions ($r \gtrsim 25 r_g$), the wind ejection rate recovers roughly the same power law as the accretion rate (see Figure 4).

If the system is in statistical steady state, the radial mass ejection power law should be related to the angular momentum power law we measured in Section 3.2,

$$\dot{L}_{\rm tot} \sim \dot{M}_{\rm disk} \langle u_\varphi \rangle, \quad (15)$$

where $\dot{L}_{\rm tot}$ is the angular momentum accretion rate, and $\langle u_\varphi \rangle$ is the averaged $\varphi$-velocity. If we assume Keplerian orbits in the accretion disk, we know that $\langle u_\varphi \rangle \propto r^{0.5}$. Therefore, we can use our radial dependence in Figure 4 to recover the radial dependence of the mass accretion rate we measure in Figure 5. Indeed, we see that $\dot{L} \propto r^{0.9}$ in Figure 4 and confirm that $\dot{M}_{\rm disk} \propto r^{0.4}$ in Figure 5, as expected. The coherence of the angular momentum and mass flux power laws is correctly predicted by self-similar theory (Ferreira & Pelletier 1995; Ferreira 1997; Jacquemin-Ide et al. 2019). The measured power law for the accretion rate, $\xi = 0.4$, is consistent with a heavily mass-loaded outflow and is in the upper bound of what can be achieved with self-similar models (Jacquemin-Ide et al. 2019). The value we measure is in contrast with Scepi et al. (2024), who measure $\xi = 1.04$, but it is consistent with McKinney et al. (2012), who measure $\xi = 0.4$.

We can see from Figure 5 that the effective value of $\alpha$ is $\gtrsim 1$ at $r \lesssim 10 r_g$, but it decreases toward unity with radius. The overall magnitude of $\alpha \sim 1$ is high, but not surprising. Wind transport, which cannot be captured by the $\alpha$ formalism, is significant and in equipartition with turbulent transport, suggesting that our effective $\alpha$ could be limited to $\lesssim 1$–2. Additionally, shearing-box MRI (Salvesen et al. 2016) and global MAD simulations (Scepi et al. 2024) both suggest that the turbulent $\alpha$ is $\propto \beta^{-1}$. Throughout our disk, we measure $\beta \sim 1$ (see Figure 1), which is consistent with an order-of-unity effective $\alpha$ (McKinney et al. 2012).

## 4. Summary

In this work, we have studied the time- and $\varphi$-averaged angular momentum transport processes occurring in a high-resolution GRMHD simulation of a thick magnetically arrested disk around a rapidly rotating ($a = 0.9$) BH, performed using the GPU-accelerated code H-AMR. Here, the MAD state has been self-consistently reached by initializing our disk with a toroidal magnetic field that develops dynamically important poloidal fields via a magnetic dynamo (Jacquemin-Ide et al. 2023). We have focused in particular on the quasi-steady-state late-time behavior of the flow by time-averaging its properties from 80,000 to 120,000 $r_g/c$, during which the $\varphi$- and time-averaged mass accretion rate remains nearly constant out to $\sim 80 r_g$ (Figure 5). Our results are listed below.

1. On long timescales, the angular momentum streamlines can be used to neatly divide the azimuthally and time-averaged flow into three regions: the jet, wind, and disk regions. We define the jet-wind boundary by iterating through the total angular momentum flux streamlines from the pole to the last streamline anchored to the event horizon; all streamlines beyond this boundary are anchored to the disk. As we move from the jet-wind boundary to lower latitudes, we encounter radially outward net angular momentum flux; we label these regions as the wind. As we approach the equator, we reach regions of radially inward total angular momentum flux; we label these regions as the disk. We then define the disk-wind boundaries as the set of turn-around points in our time- and $\varphi$-averaged streamlines of the total angular momentum flux. We determine these boundaries separately for the northern and southern hemispheres of our system, as the system shows long-term asymmetries about the equator. We label the delineated regions in Figure 2. Correctly separating these regions is essential for studying angular momentum transport mechanisms in systems where outflows are dynamically important. We also emphasize that this definition for the disk-wind boundary differs from the location of the thermal disk scale height.

2. By examining each term in the Reynolds-decomposed stress-energy tensor, we find that magnetic torques are the dominant angular momentum transport in the system at $r \lesssim 20 r_g$. We find that within the wind, advection of gas and laminar magnetic torques generally dominate (Figure 3). When we calculate the stress orthogonal to the disk-wind boundary (right panel of Figure 4), we find a similar behavior: The laminar magnetic torque dominates for $r \lesssim 20\,r_{\rm g}$, which accelerates the gas, causing the advective transport to dominate for $r \gtrsim 20\,r_{\rm g}$. The picture might be more complicated farther away from the BH, however (see Section 3.2 and the Appendix). Since magnetic and advective torques both contribute to the angular momentum transport generated by the winds, we qualify them as magneto-thermal. Within the disk, we can see that the turbulent Maxwell stress dominates the transport at all radii (left panel of Figure 4), presumably excited by some combination of the magnetorotational and magnetic interchange instabilities.

3. We compare the overall strength of the angular momentum transport in the wind versus the disk and find that they transport equal amounts of angular momentum away from the BH. In Figure 4, we can see that the laminar transport of angular momentum via the winds is remarkably close to equipartition with the turbulent transport of angular momentum occurring within the disk. Both follow nearly linear ($\propto r^{0.9}$) radial profiles, corresponding to a mass accretion rate radial profile $\dot{M} \propto r^{0.4}$ and to a specific angular momentum profile $l \propto r^{0.5}$.

4. We contextualize the radial angular momentum transport behavior within the classical $\alpha$-viscosity paradigm and confirm the mass accretion rate profile of $\dot{M} \propto r^{0.4}$. The overall rate of angular momentum transport is high, which would correspond to an effective $\alpha$ of ~1 (Figure 5). Additionally, we compare our measured mass accretion rate profile with other works, establishing agreement with McKinney et al. (2012), who measure $\propto r^{0.4}$, and disagreement with Scepi et al. (2024), who measure $\propto r^{1.04}$. Scepi et al. (2024) use the expected logarithmic derivative of the accretion rate to estimate the disk ejection index (Equation (14)), while we measure the ejection index directly from the slope of the $\dot{M}_{\rm disk}$ and $\dot{L}_{\rm tot}/\langle u_\varphi \rangle$ (see Figures 4 and 5 and Equation (15)). Our measurement is likely more robust as it relies on the consistency of our two measurements, one using $\dot{M}_{\rm disk}$ and the other $\dot{L}_{\rm tot}/\langle u_\varphi \rangle$. However, it is worth noting that the variation in the mass ejection index may arise from the fact that Scepi et al. (2024) incorporate explicit cooling in their accretion disk, while our simulations involve an adiabatic accretion disk. The mass-loading is sensitive to the heating and cooling characteristics of the accretion disk (Casse & Ferreira 2000).

Many of the accreting BHs that we observe are thought to harbor MADs, motivating us to study them more closely. This is especially true in light of the recent EHT observations of the accreting supermassive BHs in M87 (Akiyama 2019) and Sgr A* (EHT Collaboration et al. 2022), which appear to agree most with simulated MADs as compared to other accretion disk models. In fact, most low-luminosity AGN are thought to be powered by MADs, suggesting that the MAD state may be the most common accretion configuration for SMBHs (Zamaninasab et al. 2014; Nemmen & Tchekhovskoy 2015; Tchekhovskoy 2015).

In the soft state, blueshifted absorption lines, which serve as proxies for accretion disk winds, are observed in the X-ray spectra of XRBs. However, the X-ray absorption lines are absent during the hard state (Ponti et al. 2016). Our nonradiative simulations are most appropriate for modeling the hard state of X-ray binaries, which are believed to host strongly magnetized radiatively inefficient accretion flows (Ferreira et al. 2006; Remillard & McClintock 2006; Marcel et al. 2019). The absence of X-ray absorption lines during the hard state does not imply the disappearance of winds, but rather indicates changes in the ionization of the wind due to alterations in the flux of ionizing X-ray photons from the disk (Chakravorty et al. 2013; Petrucci et al. 2021). This interpretation is supported by the detection of blueshifted near-infrared absorption lines during complete state transitions, exhibiting similar properties during both the hard and soft states (Sánchez-Sierras & Muñoz-Darias 2020). Thus, accretion disk winds persist throughout the entire state transition, possibly exhibiting different properties between the hard and soft states, but not in the near-infrared. Consequently, comparing the mass-loading of our disk winds with those observed in X-ray spectra is warranted, as the extent to which the properties of the disk-wind change during the state transition remains uncertain.

Disk winds from X-ray binaries in the hard state can be modeled as Blandford & Payne (1982) accelerated outflows with field lines anchored in the accretion disk (Ferreira & Pelletier 1995). Chakravorty et al. (2016) computed the absorption lines from these outflow models and compared them to observations of X-ray binary disk winds during the soft state, finding heavily mass-loaded winds that correspond to a mass ejection index $\xi \gtrsim 0.1$ (note that $p$ is used as the mass ejection index in Chakravorty et al. 2016). More recent works constrain the mass ejection index of X-ray binaries to $\xi \simeq 0.3$–$0.4$ (Fukumura et al. 2021; Chakravorty et al. 2023). The slope we measure in our simulation, $\xi \simeq 0.4$, falls within the observed range.

Since our simulations include no radiative cooling, our results most directly apply to BHs accreting at rates far below[12] the Eddington limit, such as the hard state of X-ray binaries or low-luminosity AGN.

However, in many systems that may potentially harbor MADs, such as the system in our simulations, radiative effects are essential. For example, changing-look AGN feature rapid variability on timescales of months to years, for which strong magnetic fields have sometimes been invoked as an explanation (e.g., Dexter & Begelman 2019). However, they appear at Eddington ratios of ~0.1%–10%, where we usually expect efficiently cooling geometrically thin disks. Additionally, ultrafast outflows (UFOs), which can reach velocities up to ~$0.3c$, have also been observed in similarly subcritical AGN (e.g., Tombesi et al. 2014; Parker et al. 2017); they may be the result of magnetically driven winds. For Eddington ratios where the inner disk is dominated by radiation pressure ($\gtrsim 1\%$), classical $\alpha$-disks are thermally unstable, yet both AGN and XRBs show little evidence of thermal instability at these accretion rates. This discrepancy may be alleviated in the presence of dynamically important magnetic fields (Jiang et al. 2013; Sądowski 2016). Furthermore, accretion disk models that have dynamically important magnetic fields within the inner regions can reproduce the spectral properties of XRBs during their state transitions (Ferreira et al. 2006; Marcel et al. 2019). Therefore, it is important to study MADs with simulations that have some prescription for radiative cooling (Scepi et al. 2024), or, better yet, that include radiation transport (Liska et al. 2022a, 2023).

---

[12] Black holes accreting far above the Eddington rate also cool inefficiently, and thus may resemble polytropic models such as ours. But, at the very least, their equation of state will alter from $\gamma = 5/3$ to $\gamma = 4/3$, and the presence of radiatively driven outflows can alter the angular momentum transport mechanisms.

Scepi et al. (2024) found that when their disk was not significantly cooled,[13] laminar magnetic torques, $-\langle b^\theta \rangle \langle b_\varphi \rangle$ dominated the angular momentum transport. Furthermore, they found that the vertical advective wind torque, $\langle w u^\theta \rangle \langle u_\varphi \rangle$, was negligible. In the inner regions ($<10\ r_g$), we find a similar trend to that in Scepi et al. (2023). We also find that the laminar vertical wind torque, $-\langle b^\theta \rangle \langle b_\varphi \rangle$, is stronger by a factor of two than the turbulent angular momentum torque, $\langle \delta b^\mu \delta b_\varphi \rangle$ (see their Figure 10). In contrast with Scepi et al. (2024), we find that for $r > 20 r_g$, the total vertical laminar torque, $\langle w u^\theta \rangle \langle u_\varphi \rangle - \langle b^\theta \rangle \langle b_\varphi \rangle$, at the disk-wind boundary is in equipartition with the turbulent magnetic angular momentum transport, $-\langle \delta b^r \delta b_\varphi \rangle$, within the accretion disk (Figure 4). This discrepancy exists because Scepi et al. (2024) only measured the ratio of the torques in the inner regions of the disk, $r < 10 r_g$. Our results show that turbulent diagnostics need to be evaluated sufficiently far from the BH, $r \gtrsim 20 r_g$, for them to reach their asymptotic properties.

In Chatterjee & Narayan (2022), the authors argued that magnetic flux eruptions are a dominant mode of angular momentum transport in their MAD simulations. One major difference between the simulations, however, is that they modeled nonspinning, and therefore jetless, BHs, while our BH rotates rapidly. The importance of magnetic flux eruptions in angular momentum transport in MADs around spinning jetted BHs is yet to be studied closely. While we did not note enhanced angular momentum transport during flux eruption events, we also focused on the time- and $\varphi$-averaged angular momentum transport rather than on any time-dependent mechanisms. Furthermore, the wind region is vertically squeezed close to the BH ($r \lesssim 10 r_g$; see Figure 2) by the jet, suggesting that the pressure at the jet-wind boundary may have a strong effect on the properties of the wind outflows. It is not immediately obvious that the transport of angular momentum by winds should be the same in both jetted and jetless BHs. This would indicate that MADs harbored by spinning versus spinless (and, therefore, jetless) BHs may have different wind properties, and this merits a dedicated comparison. Clearly, these discrepancies motivate further studies of angular momentum transport in MADs.

Effective $\alpha$-models are a simplified but powerful method of modeling the secular behavior in astrophysical disks (Shakura & Sunyaev 1976; Dubus et al. 2001). However, large-scale vertical torques (advective or magnetic), which are dynamically important in our simulations and in those of Scepi et al. (2024), are difficult to model accurately with an $\alpha$-viscosity. Unlike turbulent torques, vertical laminar torques do not induce viscous spreading of the disk as they do not lead to dissipation, but only act as a sink term for angular momentum. Hence, laminar and turbulent torques lead to distinct disk behaviors on long timescales (Scepi et al. 2020; Tabone et al. 2022). Furthermore, disk winds also remove significant matter from the accretion disk (see Figure 5), completely changing their long-term impact on the disk dynamics when compared to simple $\alpha$-models. Additionally, it has become clear that the angular momentum loss via disk winds is also dynamically important for the orbital evolution of binaries that overflow their Roche lobes, such as low-mass X-ray binaries (Gallegos-Garcia et al. 2023). The ubiquity of winds in astrophysical disks necessitates updating the disk models to include disk-wind torques, as they are essential for understanding the long-term evolution of accretion disks.

### Acknowledgments

V.M. acknowledges the support of Northwestern University's Summer Undergraduate Research Grant. J.J. and A.T. acknowledge support by the NSF AST-2009884 and NASA 80NSSC21K1746 grants. A.T. also acknowledges support by NSF grants AST-2107839, AST-1815304, AST-1911080, OAC-2031997, and AST-2206471. G.M. was supported by a Canadian Institute of Theoretical Astrophysics (CITA) postdoctoral fellowship and by a Netherlands Research School for Astronomy (NOVA), Virtual Institute of Accretion (VIA) postdoctoral fellowship. Support for this work was also provided by the National Aeronautics and Space Administration through Chandra Award Number TM1-22005X issued by the Chandra X-ray Center, which is operated by the Smithsonian Astrophysical Observatory for and on behalf of the National Aeronautics Space Administration under contract NAS8-03060. This research was also made possible by NSF PRAC award no. 1615281 at the Blue Waters sustained-petascale computing project and supported in part under grant No. NSF PHY-1125915. This research used resources of the Oak Ridge Leadership Computing Facility, which is a DOE Office of Science User Facility supported under Contract DE-AC05-00OR22725.

### Appendix
### Local Surface Angular Momentum Flux

In panel (b) of Figure 4, we presented each term contributing to the integrated angular momentum flux—$G_\varphi^\theta$—leaving the disk-wind boundary (see Equation (12)). This calculation results in a cumulative integral as a function of radius. This can make it difficult to assess by eye the local radial dependence of the various terms contributing to $G_\varphi^\theta$. To remedy this, we have presented Figure 6, which is analogous to panel (b) of Figure 4, but we show $\frac{d}{dr} G_\varphi^\theta$. This quantity depicts the local vertical angular momentum escaping the surface of the disk per unit width of the boundary $dr$.

It is insightful to compare the advective surface flux ($\langle w u^\theta \rangle \langle u_\varphi \rangle$) between Figures 4(b) and 6. In particular, we saw in Figure 4 that the advective surface flux became the dominant contributor to the $r$-integrated flux at $r \sim 80 r_g$. In Figure 6, we can see that this term actually begins to dominate the laminar magnetic surface stress ($-\langle b^\theta \rangle \langle b_\varphi \rangle$) at $\sim 20 r_g$. This indicates that the outflows are not purely magnetically driven and are instead driven by a combination of magnetic and pressure forces. This is perhaps unsurprising because thermal pressure is dynamically important when the flow is adiabatic and the sound speed is approximately virial, as is the case here.

### ORCID iDs

Vikram Manikantan  https://orcid.org/0000-0003-0547-6158
Nicholas Kaaz  https://orcid.org/0000-0002-5375-8232
Jonatan Jacquemin-Ide  https://orcid.org/0000-0003-2982-0005
Gibwa Musoke  https://orcid.org/0000-0003-1984-189X

---

[13] The disk in Scepi et al. (2024) is cooled to a thermal aspect ratio of 0.3, which is near the thermal aspect ratio of our disk. As discussed above, cooling might lead to slight differences in the wind dynamics and mass-loading (Casse & Ferreira 2000). However, we do not believe this has a considerable impact on the angular momentum transport.

Koushik Chatterjee ⬤ https://orcid.org/0000-0002-2825-3590
Matthew Liska ⬤ https://orcid.org/0000-0003-4475-9345
Alexander Tchekhovskoy ⬤ https://orcid.org/0000-0002-9182-2047